\documentclass[aps,prl,reprint,superscriptaddress,showpacs,amsmath,amssymb]{revtex4-1}
\usepackage{graphicx}  
\usepackage{epsfig} 
\usepackage{subfigure}
\usepackage{lineno}  
\usepackage{amsmath,amssymb}
\usepackage{comment}
\usepackage{multirow}
\usepackage{dcolumn}% Align table columns on decimal point
\usepackage{bm}% bold math
\usepackage{setspace}
\includecomment{printrobustnotes}
\includecomment{printallnotes}
\usepackage{xspace}
\usepackage{color}
\usepackage{lineno}  

\begin{document}
%\modulolinenumbers[1]
%\linenumbers

\title{Radon variation measurements at the Yangyang underground laboratory} 

\author{C.~Ha}
\affiliation{Department of Physics, Chung-Ang University, Seoul 06974, Republic of Korea}
\author{Y.~Jeong}
\affiliation{Department of Physics, Chung-Ang University, Seoul 06974, Republic of Korea}
\author{W.~G.~Kang}
\affiliation{Center for Underground Physics, Institute for Basic Science (IBS), Daejeon 34126, Republic of Korea}
\author{J.~Kim}
\affiliation{Department of Physics, Chung-Ang University, Seoul 06974, Republic of Korea}
\author{K.~W.~Kim}
\affiliation{Center for Underground Physics, Institute for Basic Science (IBS), Daejeon 34126, Republic of Korea}
\author{S.~K.~Kim}
\affiliation{Department of Physics and Astronomy, Seoul National University, Seoul 08826, Republic of Korea}
\author{Y.~D.~Kim}
\affiliation{Center for Underground Physics, Institute for Basic Science (IBS), Daejeon 34126, Republic of Korea}
\affiliation{IBS school, University of Science and Technology (UST), Daejeon 34113, Republic of Korea}
\author{H.~S.~Lee}
\affiliation{Center for Underground Physics, Institute for Basic Science (IBS), Daejeon 34126, Republic of Korea}
\affiliation{IBS school, University of Science and Technology (UST), Daejeon 34113, Republic of Korea}
\author{M.~H.~Lee}
\affiliation{Center for Underground Physics, Institute for Basic Science (IBS), Daejeon 34126, Republic of Korea}
\affiliation{IBS school, University of Science and Technology (UST), Daejeon 34113, Republic of Korea}
\author{M.~J.~Lee}
\affiliation{Department of Physics, Sungkyunkwan University, Suwon 16419, Republic of Korea}
\author{Y.~J.~Lee}
\affiliation{Department of Physics, Chung-Ang University, Seoul 06974, Republic of Korea}
\author{K.~M.~Seo}
\affiliation{Department of Physics and Astronomy, Sejong University, Seoul 05006, Republic of Korea}

\date{\today}

\begin{abstract}
  The concentration of radon in the air has been measured in the 700~m-deep Yangyang underground
  laboratory between October 2004 and May 2022. 
  The average concentrations in two experimental areas, called A6 and A5, were measured
  to be 53.4$\pm$0.2~$\rm Bq/m^3$ and 33.5$\pm$0.1~$\rm Bq/m^3$, respectively.
  The lower value in the A5 area reflects the presence of better temperature control and ventilation.
  The radon concentrations sampled within the two A5 experimental rooms' air are found to be correlated to the local surface temperature outside of the rooms, with correlation coefficients $\rm r = 0.22$ and $\rm r = 0.70$. Therefore, the radon concentrations display a seasonal variation, because the
  local temperature driven by the overground season influences air ventilation in the experimental areas.
  A fit on the annual residual concentrations finds that the amplitude occurs each year on
  August, 31 $\pm$ 6 days.
\end{abstract}

\maketitle

\section{Introduction}
\label{sec:intro}
The materials comprising our universe are predominantly radiationless dark components
of a nature that is not adequately understood.  Based on astrophysical observations, it has been
determined that 26\% of all energy is formed by this so-called dark matter~\cite{Planck:2018vyg}.
Theoretically, the composition of dark matter often modeled as various particles beyond the Standard
Model~\cite{1937ApJ86217Z,Workman:2022ynf}, wherein a weakly interacting massive particle (WIMP)
is one of the most frequently considered candidates~\cite{lee77}. Searches for  WIMPs are being
conducted by a number of groups using a variety of experimental
approaches~\cite{Schumann:2019eaa,Gaskins:2016cha,Kahlhoefer:2017dnp,XENON:2022mpc,LZ:2022ufs}. One of these involve attempts
to measure the energy deposited by nuclear recoils from WIMP interactions in the material
of low-background detectors.

To date, no unambiguous evidence for WIMP--nucleus interactions has been reported, other than an
annual modulation of residual events that cannot be explained by known background sources and
has been interpreted by some authors as a signal of the yearly changes in the direction of the
Earth’s orbital motion in a galactic halo of dark-matter WIMPs~\cite{Bernabei:2018yyw,Freese:2012xd}.
On the other hand, other, more mundane sources of modulations, such as cosmic-ray muon induced
reactions, or radon concentrations in the air of the laboratory environment have been suggested and
studied~\cite{Davis:2014cja,Blum:2011jf,Tiwari:2017xnj,MINOS:2012xaa,McKinsey:2018xdb}.
As part of these studies, measurements of the time dependence of the radon concentration are essential.

Radon is produced as a daughter nuclei decay product from the radioactive nuclides in the material of the
surrounding tunnel walls.  Specifically, this material contains traces of uranium and thorium that
are the primary sources of radon.
For example, $^{222}$Rn (t$_{1/2}$=3.82 days) is generated from the $^{238}$U decay chain and,
being an inert gas, it can deposit onto the surfaces of detector materials.
Its decay products such as $^{210}$Pb (t$_{1/2}$=22.2~years) and $^{210}$Bi (t$_{1/2}$=5.0~days) decay by emitting a beta particle which can act as background sources at low energies while the other daughter elements like $^{210}$Po (t$_{1/2}$=138.4~days) produces an alpha particle and a nuclear recoil which can affect region of interest in rare decay experiments.
Also, when the radon decays into its daughter isotopes, several gamma rays are
produced and can contribute to the background spectrum of the dark matter data.
Much effort has been made to understand the radon production mechanism and its mitigation in rare decay experiments
especially at the underground laboratory environment~\cite{Pronost:2018ghn,Liu:2018abg,Perez-Perez:2021bzu,Hod_k_2019,Murra:2022mlr,LZ:2020fty}.

The Yangyang underground laboratory hosts two dark matter experiments and one neutrinoless double-beta
decay experiment and, since 2004, the underground radon level with both
custom-designed and commercially available detectors has been monitored.
In this note, we report on the long-term variations in the radon level that is
based on an analysis of these measurements.

\section{Materials and Methods}

\subsection{Experimental sites at Yangyang underground laboratory}

The Yangyang underground laboratory (Y2L) is located adjacent to the underground generator of the
Yangyang pumped water plant in east Korea. The plant contains the main access tunnel with auxiliary
tunnels, named as A5 and A6, that house the experimental facilities. Fresh air from the surface enters
the tunnels through the main ramp way and is pumped out via a separate duct.
The power plant operates at minimum two exhaust fans continuously, each with 63000 m$^3$/h capacity.
This exchanges the entire air volume of the 2.2~km-long main tunnel and auxiliary spaces
once every 40 minutes.
Throughout the year, the temperature inside the tunnel is maintained between 22~$^{\circ}$C and 26~$^{\circ}$C, and the relative
humidity in the areas surrounding the laboratory is kept in the ranges of 60--70\%. The minimum granite
overburden in these areas is 700 m (1800 meter-water-equivalent depth) and the cosmic-ray muon fluxes at A5 and A6 (two are situated 300~m
apart horizontally) are measured to be
$\rm 3.795\pm0.110\times 10^{-7}~s^{-1}~cm^{-2}$~\cite{Prihtiadi:2017inr,COSINE-100:2020jml} and
$\rm 4.4\pm0.3\times 10^{-7}~s^{-1}~cm^{-2}$~\cite{KIMSmuonflux}, respectively. The subterranean rock is
primarily composed of gneiss that contains 2.1~ppm and 13.0~ppm of uranium and thorium, respectively,
as measured by an inductively coupled plasma mass spectrometry~\cite{Lee:2011jkps}.

The Korea Invisible Mass Search (KIMS) experiment~\cite{Kims:2005dol} in the A6 tunnel operated a
CsI (Tl) crystal array to search for dark matter for over 15 years, and it is currently used
for R\&D activities associated with the development of ultralow-background crystals.
The COSINE-100 experiment~\cite{Adhikari:2017esn,Adhikari:2018ljm} uses NaI(Tl) crystals and is
currently operating in the A5 experimental space. In addition, other experiments, including a search
for neutrinoless double-beta decay experiment (called AMoRE-I)~\cite{Alenkov:2019jis} and a 
high-purity Germanium array (HPGe)~\cite{Leonard:2020izl} are operating in other rooms situated
in the A5 tunnel. The A5 area is equipped with a radon reduction system (RRS) that supplies
radon-filtered air to each experimental room as required.
While the RRS is providing radon-reduced
air with a flow of 140 m$^3$/h to an experimental room, the measured radon level is  10~to~100 times lower than that of the
room air~(Since 2015, Ateko system (www.ateko.cz) has been operating at A5 and the specified reduction is a factor 1000 or more.).

\begin{figure*}[htbp]
  \begin{center}
    \includegraphics[width=0.95\textwidth]{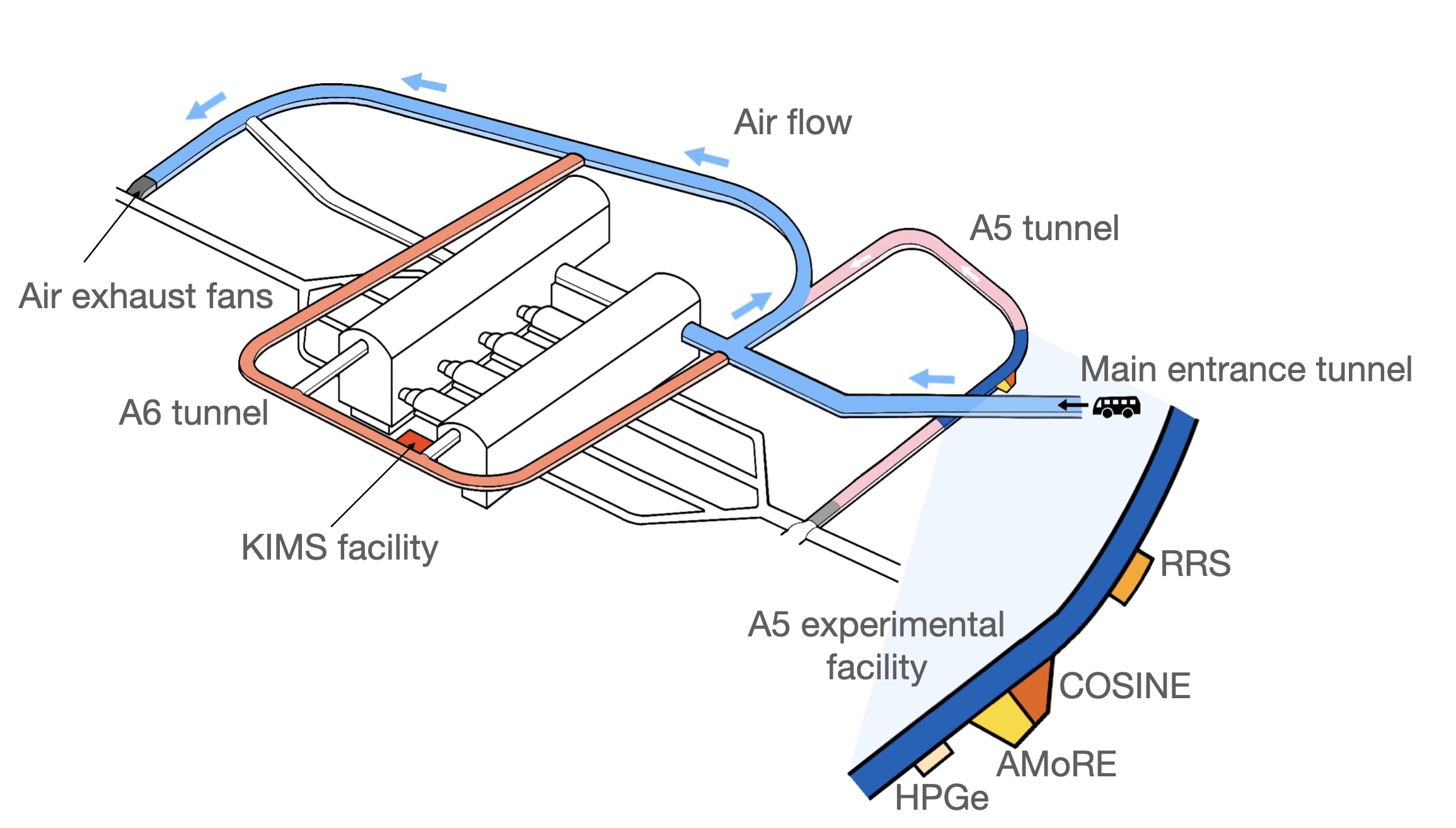}
    \caption{A map of Y2L.
      The experimental areas are accessed by vehicles through the main entrance.
      Since air is exhausted by fans at the end of the main tunnel, fresh air flows
      in a single direction.
      The KIMS experiment was located in the A6 tunnel, whereas the newer facilities
      are situated in the A5 tunnel, including the COSINE-100, AMoRE-I, and HPGe experiments.
      A5 and A6 are horizontally separated by 300~m, and A5 is approximately 50~m deeper than A6.
    }
    \label{orient}
  \end{center}
\end{figure*}

Since the A6 tunnel area is separated from the main tunnel by doors, the  air flow rate
is somewhat restricted.  In contrast, the entrance of the A5 tunnel is always open and the tunnel is
equipped with its own air supply system that provides relatively robust air circulation.
The system supplies the main tunnel air to the end of the A5 tunnel using two 2300~m$^3$/h fans
which exchange the branch tunnel air once every hour.
A schematic drawing of the Y2L arrangement with the locations of the A5 and A6 areas indicated
is provided in Fig.~\ref{orient}.

The COSINE-100 experiment is housed in an environmentally regulated room with controlled humidity
and temperature.  Its detection room has a floor area of $\rm 44~m^2$ and a ceiling height of $\rm 4~m$. 
The air control system maintains the room temperature at $\rm 23.5\pm0.1~^{\circ}C$ and relative humidity
at $37\pm1~\%$.  The air in the room is continuously circulated through a HEPA filter, and the number
of dust particles larger than $\rm 0.5~\mu m$ is maintained below 1500 per cubic foot.
The room air is sampled by a radon detector.
These environmental parameters in the experimental room and in the tunnel are monitored online.
The details of experimental control are described in~\cite{COSINE-100:2021mlj}.

\subsection{Radon Counter Setup}
\begin{figure*}[htbp]
  \begin{center}
    \includegraphics[width=0.99\textwidth]{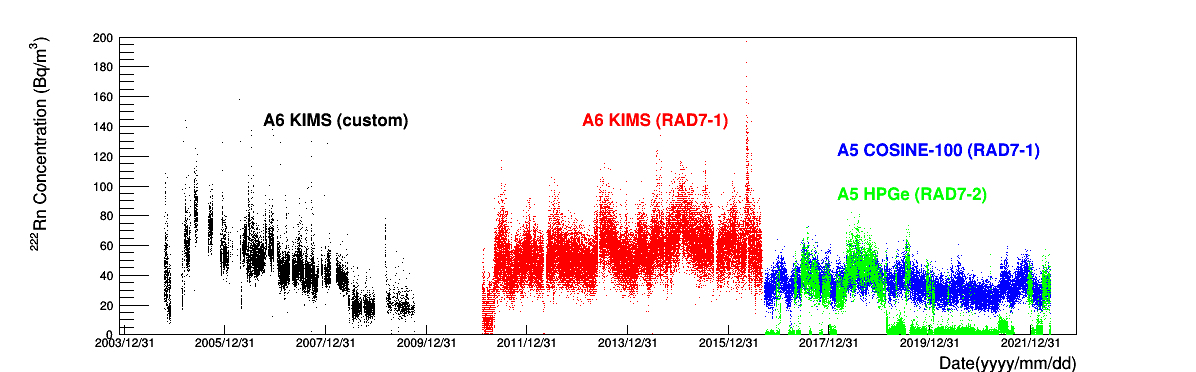}
    \caption{The $^{222}$Rn concentration in Y2L between 2004 and 2022 was measured in two different
      experimental areas. At the KIMS laboratory, the custom detector measurements were recorded
      between October 2004 to October 2009 (black), whereas the RAD7-1 measurements were recorded
      between February 2011 and September 2016 (red).
      The COSINE-100 room data were acquired between September 2016 and May 2022 (blue) with the
      same RAD7-1 counter and the HPGe room data (RAD7-2) are displayed in green marks. 
    }
    \label{rad7}
  \end{center}
\end{figure*}

Since 2004, a custom-design radon detector measured the radon concentration at the A6 KIMS detector room.
The detector consisted of 69.3 liters of a chamber for air sampling and a 900~mm$^2$ silicon PIN diode
for an alpha particle sensor on which an ionized Polonium isotope can attach and decay~\cite{Lee:2011jkps}.
In 2011, this detector was replaced with a commercially available detector from Durridge company~(RAD7-1)~\cite{duridge}.
In 2016, the RAD7-1 detector was moved to the COSINE-100 detector room, where it has been functioning
ever since.  Another commercial counter~(RAD7-2, the same model as RAD7-1) was installed in the HPGe detector room in 2016. In
the RAD7 devices, a silicon diode sensor is located at the center a drift chamber with an applied
electric field.  When a $^{222}$Rn nucleus decays inside the drift chamber, it produces a positively
charged $^{218}$Po ion that drifts to and sticks on the surface of the diode's sensitive area.
Within a few minutes, the $^{218}$Po nucleus decays into a $^{214}$Pb nuclei and an alpha particle,
and the alpha particle produces an energy in the diode. The rate per unit volume for these signals reflects the
mother $^{222}$Rn isotope concentration in the drift chamber air volume.
The specified RAD7 detection limit is at 4~Bq/m$^3$.
In this way, the total radon
level in the room air is  measured every two hours and the recorded data are transmitted to a slow
monitoring server, as displayed in Fig.~\ref{rad7}.

The RAD7 detectors have been cross-calibrated at various locations and cross-checked
with a separate custom-made detector, including a commercial ion chamber detector (RadonEye~\cite{radeye}).
All these tests yielded consistent results and
the RAD7 detectors did not exhibit any abnormal behavior as long as
the desiccants were regularly replaced to maintain a stable humidity-level inside the chamber.
This new custom detector is the same type and the same dimension as the original A6 KIMS radon counter and is under testing and development phase. The 70~L chamber is made of stainless steel and the sensor uses Hamamatsu Si PIN photodiode (S3204-9, $18\times18$ mm$^2$), HV divider, and preamplifier (Hamamatsu H4083).
When the RRS output air is directly connected to this custom radon detector,
the concentration was measured to be 26.1$\pm$5.6 mBq/m$^3$ which is more than a factor of 1000 reduction.

\subsection{Data collection}

During the 4762 days of the total operating period between October 2004 and May 2022, radon data
were acquired in three distinct periods. At the A6 lab, The KIMS custom-made detector operated
for $\sim$5 years until October 2009.  Those data are reported in Ref.~\cite{Lee:2011jkps}, for which
a prescale factor of 10 was applied. After a 16 month period  of no measurements, the RAD7-1 detector
was installed at the same location.  In September 2016, the RAD7-1 detector was moved to the A5
COSINE-100 room for data acquisition.
In a similar timeframe, we operated another detector---RAD7-2 in the HPGe room at A5. Overall, the
RAD7 detectors were operated continuously with a short dead times that were primarily caused by
power outages in the tunnel.  The analyses reported here use all of the acquired data.

\begin{figure}[htbp]
  \begin{center}
    \includegraphics[width=0.48\textwidth]{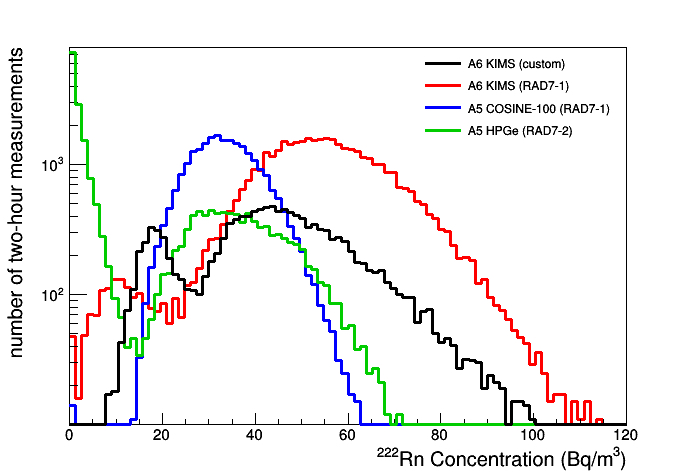}
    \caption{The distributions of $^{222}$Rn concentrations in the different Y2L detector rooms.
      Radon levels were compared among three distinct Y2L experimental areas.
      A Gaussian fit is performed on the A6 KIMS lab measurement (red) that has a peak value of
      $53.4\pm0.2$~$\rm Bq/m^{3}$ with a width of $13.9\pm0.3$~$\rm Bq/m^{3}$,
      The measurements for the A5 COSINE-100 laboratory (blue) have the fit mean of $33.5\pm0.1$~$\rm Bq/m^{3}$
      and fit width of $7.9\pm0.1$~$\rm Bq/m^{3}$.
      The measurement on the A5 HPGe room (green) exhibits two peaks corresponding to measurements
      with and without the supply of radon-reduced air.
      Note that A6 KIMS (custom) measurements also have two peaks and the lower peak is due to
      different ventillation condition executed in the period between middle of 2008 and late 2009.
    }
    \label{radvar}
  \end{center}
\end{figure}

The $^{222}$Rn concentration (in Becquerel per cubic meter) is displayed in Figure~\ref{rad7} as a function of the
date for all acquired data. In particular, four distinct measurements were acquired for roughly
five years and are correspondingly color-coded.
As listed in Table~\ref{types}, the average concentrations were at the level of 1~pCi/L~(=37~Bq/m$^3$) at A5, which is
relatively low compared to  measured levels at other underground labs without RRS operating~\cite{Pronost:2018ghn,Liu:2018abg,Perez-Perez:2021bzu,Hod_k_2019}.
%The long-term structure in the concentration variation primarily results from the air circulation in the tunnel.
From late 2008 to late 2010, the temporal variations that occurred in the A6 ventilation system provided
increased airflow in that area, which caused lower concentration in that period.
The fundamental reason for the occasional short-term spikes in the data is because periods of increased 
caused by inadequate maintenance of the chamber air desiccant.
The detector specified 5\,\% accuracy at normal humidity levels~\cite{duridge}.
The distributions of radon concentrations were fitted with a Gaussian function
and the fit means and sigmas are obtained.
Their statistical uncertainties are obtained as one unit increase from minimum $\chi^2$ value
by a parameter scan against $\chi^2$ values.
The distributions are illustrated in Fig.~\ref{radvar}.

\begin{figure*}[htbp]
  \begin{center}
    \includegraphics[width=0.95\textwidth]{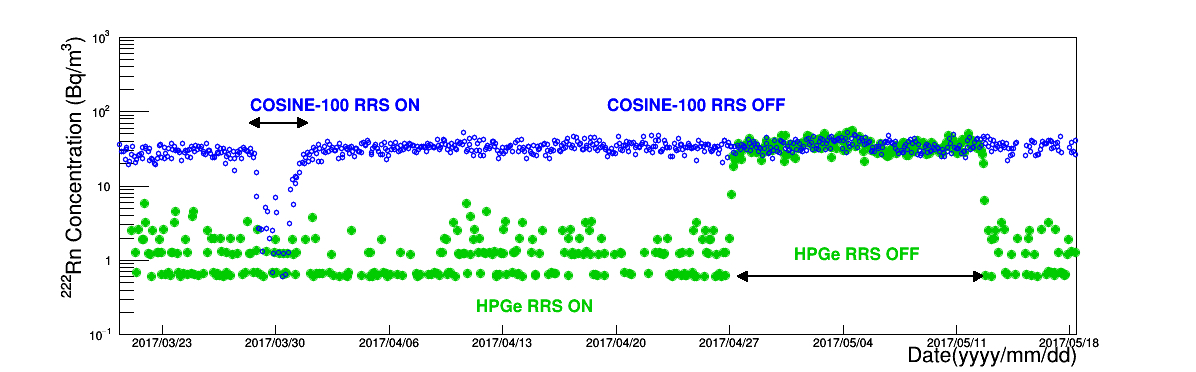}
    \caption{Comparison of two simultaneous A5 measurements of $^{222}$Rn concentrations.
      A two-month period comparison of COSINE-100 (blue circles) and HPGe (green dots) measurements is shown.
      When RRS air is supplied for COSINE-100 (RAD7-1) at around 2017/03/30, the $^{222}$Rn level drops as low as 0.6 Bq/m$^3$.
      For HPGe (RAD7-2), when RRS air is supplied, the level is on average 1.3 Bq/m$^3$ with standard deviation of 1.0 Bq/m$^3$.
      When RRS is stopped for two weeks starting from 2017/04/27,
      the measured concentration reaches to the level of 34.2$\pm$0.6~Bq/m$^3$ which is consistent with the COSINE-100 room measurement.
    }
    \label{rad712}
  \end{center}
\end{figure*}

In the case of supplying radon-reduced air into one of the experimental rooms, the radon level
sampled by the RAD7 counters is typically reduced to a few Bq/m$^3$.
The RRS air was supplied for a short period (less than a week) in the COSINE-100 room
when the NaI(Tl) crystal detector installation
and special maintenance occur.
For the HPGe room and the AMoRE-I room, the RRS air is supplied for more often and longer time span.
When the RRS air is supplied for the COSINE-100 detector room, the $^{222}$Rn level drops as low as 0.6 Bq/m$^3$
and similarly for the HPGe room.
When RRS is stopped, the measured concentration in RAD7-2 (HPGe) show
the consistent results as the RAD-1 (COSINE-100) measurements shown in Fig.~\ref{rad712}.

The measurement locations, detector type, periods, and measured radon concentrations are summarized in
Table~\ref{types}.
\begin{table*}[ht]
  \begin{center}
    \caption{
      The detectors and their locations are listed for all of the acquired Y2L radon data.
      The KIMS-Custom measurement data used a prescale factor of 10,
      and the radon-level values include systematic uncertainties; those for the other measurements
      are only statistical uncertainties.
      The radon level in HPGe measurement represents values when the RRS air is not
      being supplied ($^{*}$).
    }
  \label{types}
  \begin{tabular}{lccc}
    \hline                                                                                
    Counter Type   & Period  & Fit Mean (Bq/m$^3$)[ Fit Sigma ] & Live days \\
    \hline                                                                                
       A6-KIMS Custom   & 2004.10.18--2009.10.05  & $44.4\pm18.1$       [ $11.7\pm4.8$]&847 \\
       A6-KIMS RAD7-1   & 2011.02.14--2016.09.01  & $53.4\pm0.2$        [ $13.9\pm0.3$]&1872\\
       A5-COSINE-100 RAD7-1   & 2016.09.23--2022.05.27  & $33.5\pm0.1$  [ $7.9\pm0.1$]& 2043\\
       A5-HPGe RAD7-2   & 2016.09.28--2022.05.27  & $35.2\pm0.2^{*}$     [$10.8\pm0.2^{*}$] &1845\\
    \hline  
  \end{tabular}
  
  \end{center}
\end{table*}

\section{Results}
With the acquired data, we performed analyses among the measurements in terms of their daily concentrations and
as a function of time.  The daily concentration measured at A5 was 33.5$\pm$0.1 Bq/m$^3$, which was less than
that of A6 by 37\%. The A6 tunnel is a
both-end closed space with minimal airflow, whereas the entrance of the A5 tunnel is one-ended,
with air circulation that is superior to that in A5.
Additionally, the spread of the measurements is 7.9$\pm$0.1~Bq/m$^3$, and much less than
13.9$\pm$0.3 Bq/m$^3$ for A6.
\begin{figure*}[!htb]
  \begin{center}
      \includegraphics[width=0.9\textwidth]{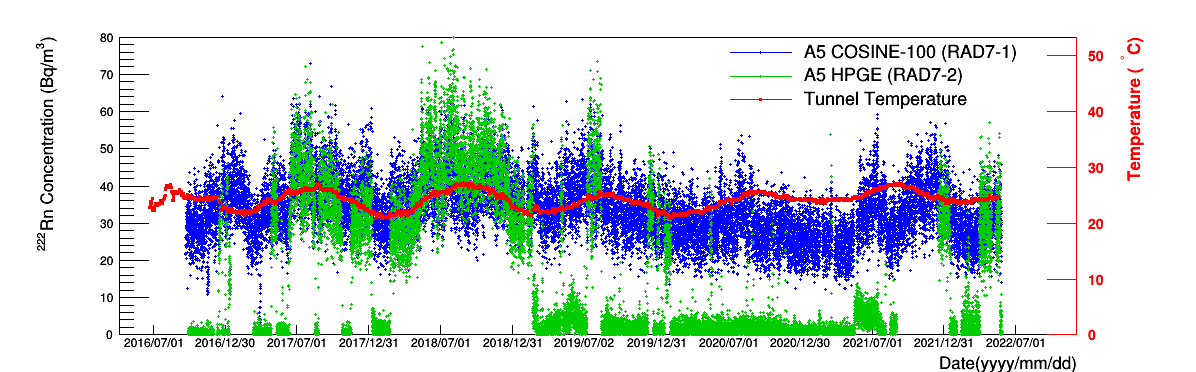}
  \end{center}
  \caption{The $^{222}$Rn concentration in the
    HPGe room (green) and the COSINE-100 (blue), which are separated by a distance of approximately
    35~m, are compared.
    The two detector rooms are monitored by the same model (RAD7) detector.
    The occasional reduction in the radon level at HPGe was caused by the occasional supply of
    radon-free air to its detector room.
    For days when the RRS was off, two detector room measurements were correlated, implying that
    the fundamental radon activity is the same in both places.
    The tunnel temperature (red), independently measured outside of the detector rooms and located in between two rooms,
    is displayed with a right ordinate label.
  }
  \label{rnhpge}
\end{figure*}

Correlations of long-term variations in the A5 measurements with
the tunnel temperatures have  also been investigated.
The concentrations measured at the A5 COSINE-100 and HPGe rooms were compared to each other as well as 
with the temperature in the A5 tunnel. The two experimental rooms are separated
by a distance of approximately 35 m.  When the RRS was not operating, the comparative variations in
radon levels were strongly correlated, as depicted in Fig.~\ref{rnhpge}.

Since the power plant company operates air circulation fans at the end of the main tunnel throughout
the year, the tunnel temperature is correlated with the surface temperatures in the immediate vicinity.
The temperature measured at A5 shows an annual variation between 22~$\rm ^{\circ}C$ and 26~$\rm ^{\circ}C$
(Fig.~\ref{rnhpge}). In the Yangyang region, the temperature averaged over the year is 11.8~$\rm ^{\circ}C$,
with an average minimum of $\rm -2.2~^{\circ}C$ in January and an average maximum of 24.3~$\rm ^{\circ}C$
in August.

\begin{figure}[!htb]
  \begin{center}
      \includegraphics[width=0.48\textwidth]{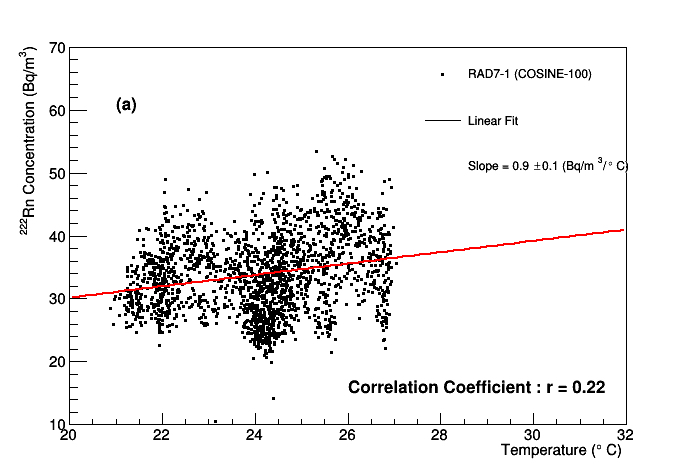}
      \includegraphics[width=0.48\textwidth]{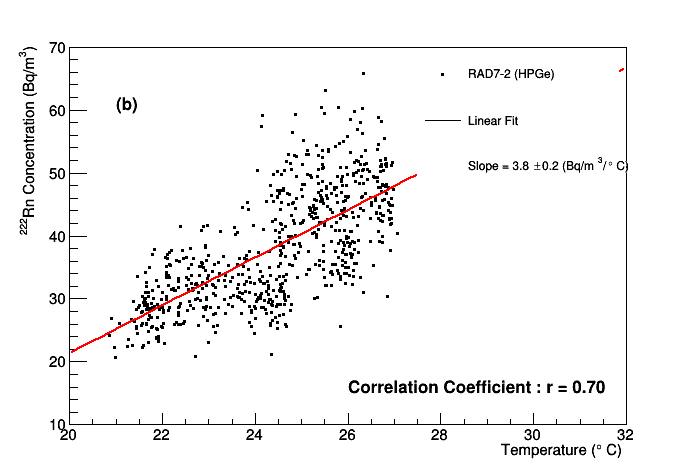}
  \end{center}
  \caption{The $^{222}$Rn concentrations versus the tunnel temperature.
    The radon concentrations for the COSINE-100 room (a) and the HPGe room (b) are
    plotted against the temperature. A linear fit have a
    slope of 0.9$\pm$0.1 Bq/m$^{3}/^\circ$C and the Pearson correlation coefficient is r = 0.22 for the RAD7-1 counter.
    For RAD7-2, the slope of 3.8$\pm$0.2 Bq/m$^{3}/^\circ$C and the coefficient of r = 0.70 are measured.
    Note that RAD7-2 measurements include data when RRS is off. 
  }
  \label{corr}
\end{figure}

The $^{222}$Rn concentrations measured in the COSINE-100 and HPGe rooms are compared with the temperature inside the
A5 tunnel in Fig.~\ref{corr}. The A5 COSINE-100 (RAD7-1) radon levels correlate with the tunnel temperature
with a correlation coefficient of $\rm r = 0.22$; the slope of a linear fit to the data was 0.9$\pm$0.1 Bq/m$^{3}/^\circ$C.
For the A5 HPGe (RAD7-2) levels, when RRS is off, the larger coefficient of $\rm r = 0.70$ is obtained
and the slope parameter was 3.8$\pm$0.2 Bq/m$^{3}/^\circ$C.

For the annual variation study, we applied an additional selection criterion to all of the acquired
KIMS and COSINE-100 radon data that eliminated all the data prior to 2011/05/11, for which the knowledge
of the ventilation conditions in the A6 tunnel was incomplete.
The RAD7-2 (HPGe) data is not used in this analysis because the RRS-off data spans relatively short periods.
The combined data period is from 2011/05/11 to 2022/05/27 (4034 days in total) and the final analysis sample
contains 3822 live days, which is 95\% of all the days in this period.
Here, we treat two-hour RAD7 measurement as a single data point, and
each daily measurement is the statistical average of the 12, two-hour measurements on that day.
The daily averages were further combined into eight-day averages.

\begin{figure*}[!htb]
  \begin{center}
      \includegraphics[width=0.98\textwidth]{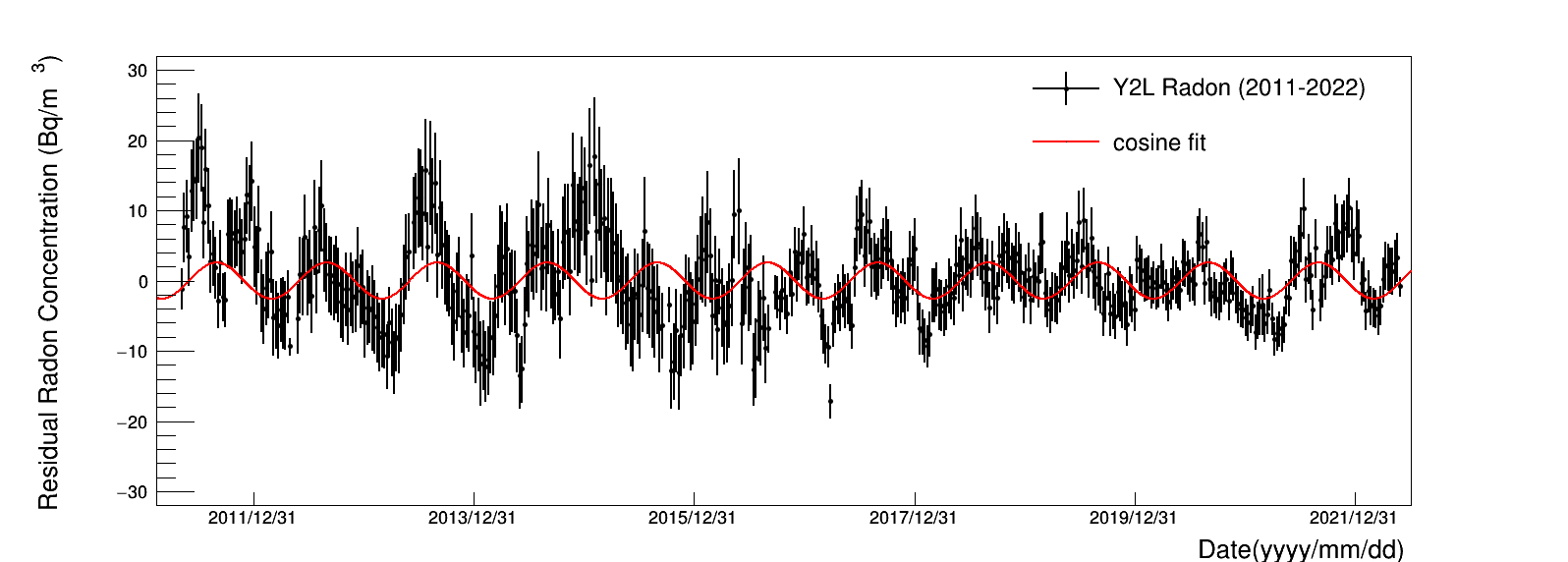}
  \end{center}
  \caption{The KIMS and COSINE-100 rooms' $^{222}$Rn concentration residuals as a function of time are
    fit with Eq.~\ref{eq}.
  }
  \label{rnresidual}
\end{figure*}

Initially, we evaluated an annual average using a period of 365.25~days, with January 1$^{st}$ as the starting time.
After subtracting the average values, the residual concentrations for each year were obtained and combined
for the entire analysis period.
Then we applied a cosine fit,
\begin{equation} \label{eq}
  f(t)=A\cos{\bigg(\frac{2\pi}{365.25}( t - t_0)\bigg)}
\end{equation}
to these residuals. 
In the Eq.~\ref{eq} fit, the period was fixed at 1 year and we fit for the amplitude $A$ and phase $t_0$ using
the $\chi^2$ method.
The best-fit $A$ was $-2.57\pm0.25$~Bq/m$^3$ and the best-fit phase at the amplitude was $60.6\pm5.6$ days with $\chi^2/NDF=742.7/497$. This corresponds to August $31\pm 6$~days at the positive maximum amplitude.
A constant linefit on the same data assuming no modulation shows $\chi^2/NDF=813.68/498$.
With $\chi^2$ difference of 71.6 between two hypotheses,
therefore, no modulation hypothesis is disfavored at more than 8 $\sigma$.
The radon-concentration residuals and the results of the fit are displayed in Fig.~\ref{rnresidual}.

For a consistency check, we performed the same fit on residual data
that combine A5 and A6 data with only their average concentrations subtracted
rather than subtracting the yearly residuals.
This fit finds an amplitude of $-2.29\pm0.25$~Bq/m$^3$ and a phase of $55.3\pm6.2$ days,
which is consistent with the yearly residual fit.
The relatively large $\chi^2$ value from the modulation best-fit has been investigated.
We evaluated that the main cause is difference in the residual fluctuations between A5 and A6.
The A6 measurements show a larger spread in the $^{222}$Rn measurement than that of A5.
The cosine fit with only the A5 yearly residuals returns an amplitude of $-2.55\pm0.29$~Bq/m$^3$ and a phase of $55.7\pm6.6$ days with improved $\chi^2/NDF$ = 346.1/258.
On the other hand, the fit with A6 only data shows worse constraints with an amplitude of $-2.67\pm0.48$~Bq/m$^3$ and a phase of $73.1\pm10.6$ days with $\chi^2/NDF$ = 394.0/237.
Additionally, the tunnel temperature is affected by other local impacts that can influence
the radon measurements. These include occasional power outages, malfunction of branch tunnel's ventilation system,
and excess heat generated by other operating experiments in the same tunnel.

\section{Discussion}
We fitted the A5 temperature annual variation with the Eq.~\ref{eq} function plus a constant term
using the least square method. We found a phase of 58.4$\pm$5.2~days and an amplitude of $-1.7\pm0.2~^{\circ}$C.
The best-fit phase is consistent with the best-fit phase of the radon concentration variations.
The seasonal variation of the radon concentration has been studied in other places with various models~\cite{2017IJMPA..3243004W,Arvela}.
They consider that variations of water contents in local soil facilitate radon diffusion when season changes.
However, in the Yangyang area, measurements of local indoor $^{222}$Rn and soil $^{226}$Ra concentrations show weak correlations~\cite{yjkim1}. Located not far from the sea, the area is relatively windy with an average wind speed of about 2.0~m/s, providing fresh air throughout the year. 
Moreover, reports show a seasonal variation of indoor radon concentration in Korea
is the highest in winter and the lowest in summer~\cite{ckkim1}, which is opposite of the water diffusion model
and what is observed in Y2L.
On the other hand, the temperature dependent radon emanation from the surrounding rocks of Y2L could influence
the observed radon concentration modulation. However, the temperature variation of the A5 branch tunnel
is $\pm2.0^{\circ}~$C from the average value.
Therefore, the effect due to the small temperature change may not be the magnitude that we observe in this study~\cite{Sakoda}.
Thus, we concluded that the observed radon modulation results from the temperature changes in
the air caused by the ventilation system, which drives the temperature variations in the tunnel.

When the main tunnel draws warm air from outside in the summer,
the air circulation in the A5 and A6 branch tunnels deteriorates because of the weak temperature
gradients between the main and the branch tunnels.
Moreover, the air conditioning fans, water cooling systems, and RRS generate
more exhaust heat in the summer time at the branch tunnels. 
Conversely, in the winter, the temperature gradient is higher and the air exchanges more rapidly.

These phase of the radon concentration modulation (Maximum on August 31$^{st}$) lags
that of the DAMA/LIBRA signal ({\it i.e.}, June 1$^{st}$).
It also lags that of the measured modulation of the cosmic-ray muon
rate in the COSINE-100 (June 27$^{th}$) by about two months. At the moment, 
the results of the COSINE-100~\cite{PhysRevD.106.052005} and ANAIS~\cite{PhysRevD.103.102005}
signals for an annual modulation in the recoil nucleus event rate
are statistics-limited and, therefore, cannot be directly compared to the radon concentration and muon rate variations.

The $^{222}$Rn concentration in the air has been measured over the past 18 years in the Y2L laboratory.
The average concentration is 53.4$\pm$0.2~Bq/m$^3$ in the A6 laboratory and 33.5$\pm$0.1~Bq/m$^3$ in the A5
laboratory; the latter is lower by 37\% because of better temperature-control and ventilation systems.
In this analytical study, we determined that the radon concentration is correlated to the tunnel temperature.
The COSINE-100 room radon concentration and tunnel temperature are correlated with the coefficient of r=0.22
while the coefficient between the HPGe radon concentration and tunnel temperature is r=0.70.
With the selected data, the yearly residual data were fit with a cosine function and the phase of the
maximum amplitude was determined to be August 31$\pm$6 days, which coincides with the phase of temperature
variations in the adjoining tunnel.
Overall, this is one of the longest running measurements of the radon concentration in underground laboratories.

\section*{Acknowledgments}
We thank the Korea Hydro and Nuclear Power~(KHNP) Company for providing the underground laboratory space at Yangyang.
This research was supported by the Chung-Ang University Graduate Research Scholarship in 2022 and
by the National Research Foundation of Korea~(NRF) grant funded by the Korean government~(MSIT) (No. 2021R1A2C1013761).

\end{document}